\documentclass{iau}
\usepackage{graphicx}

\title[Follow-up observations of Planck cold clumps in $^{12}$CO/$^{13}$CO/C$^{18}$O (1--0) transitions]
{Follow-up observations of Planck cold clumps in $^{12}$CO/$^{13}$CO/C$^{18}$O (1--0) transitions}

\author[Yuefang Wu, Tie Liu,  \& Fanyi Meng]   
{Yuefang Wu$^{1}$, Tie Liu$^{1}$,  \& Fanyi Meng$^{1}$}

\affiliation{$^1$Department of Astronomy, Peking University, \\ Postbus 100871,
Beijing, China \\ email: {\tt yfwu.pku@gmail.com} }

\pubyear{2012}
\volume{292}  
\jname{Molecular Gas, Dust, and Star Formation in Galaxies}
\editors{Tony Wong \& Juergen Ott Editor, eds.}
\begin{document}

\maketitle

\begin{abstract}
The Planck Early Cold Cores Catalog (ECC) provides an unbiased list of Galactic cold clumps, which form an
ideal sample for studying the early phases of star formation (\cite[Planck Collabrators et al.  2011]{Planck_etal11}). To study their properties, we have carried out a molecular line ($^{12}$CO/$^{13}$CO/C$^{18}$O) survey towards 674 Planck cold clumps in the ECC with the PMO 13.7 m telescope.

\end{abstract}

What is the origin of the initial mass function (IMF) of new born stars and the star formation efficiency? What is the mode of star formation (clustered vs. isolated)? How do these properties of star formation depend on the initial conditions of prestellar clumps? To make progress in the understanding of star formation, the physical properties of pre-stellar cores need to be investigated in a variety of samples. Unfortunately, the properties of pre-stellar cores are not well-known due to a lack of an appropriate large sample prior to the Planck satellite. As a by-product of Planck satellite, the so called Planck cold clumps have typical temperature of $<$14 K and H$_{2}$ column density of $\sim$10$^{22}$ cm$^{-2}$, most of which are pre-stellar clump candidates.

We have surveyed 674 Planck cold clumps in the J=1-0 transitions of $^{12}$CO/$^{13}$CO/C$^{18}$O with the PMO 13.7 m telescope (\cite[Wu et al.  2012]{Wu_etal12}). So far, more than 400 of them have been mapped and the mapping project is still on-going. All the cold clumps were detected with $^{12}$CO and $^{13}$CO emission, and 68\% of them have C$^{18}$O emission. Kinematic distances are estimated from the velocities of $^{13}$CO lines, and half of the clumps
are located within 0.5 and 1.5 kpc. We found most of these Planck cold clumps are turbulence dominated. Planck cold clumps are the most quiescent among the samples of weak red IRAS, infrared dark clouds, UC H{\sc ii} candidates, extended green objects, and methanol maser sources, suggesting
that Planck cold clumps have expanded the horizon of cold astronomy. By analyzing the
distributions of the physical parameters of the Planck cold clumps in Orion complex, we suggest that turbulent flows can shape the clump structure and dominate
their density distribution in large scale (GMC scale), but not function in small scale ( Clump scale ) due to local fluctuations (\cite[Liu et al.  2012]{liu_etal12}). The dense cores in the Planck clumps are most likely found to be gravitationally bounded rather than pressure confined. The lognormal behavior of the core mass distribution is most likely determined by internal turbulence (\cite[Liu et al.  2012]{liu_etal12}).

\end{document}